\RequirePackage[l2tabu,orthodox]{nag}
\documentclass[10pt]{eptcs}
\pdfoutput=1

\usepackage[T1]{fontenc}
\usepackage{lmodern}
\usepackage{fixltx2e}
\usepackage{microtype}
\usepackage{xspace}
\usepackage{enumitem}

\newcommand{\figureline}{\rule{\textwidth}{0.5pt}}

\makeatletter
\newcommand\etc{etc\@ifnextchar.{}{.\@}\xspace}

\makeatother

\usepackage{graphicx}

\newcommand{\inlinegraphic}[2]{
  \dimendef\grafheight=255\dimendef\grafvshift=254
  \grafheight=#1
  \grafvshift=-0.5\grafheight
  \advance\grafvshift by 0.5ex
  \raisebox{\grafvshift}{\includegraphics[height=\grafheight]{images/#2}\xspace}
}

\newcommand{\ninlinegraphic}[2][1.0]{
  \dimendef\grafheight=255\dimendef\grafvshift=254
  \setbox0 = \hbox{\scalebox{#1}{\includegraphics{images/#2}}}
  \grafheight=\the\ht0
  \grafvshift=-0.5\grafheight
  \advance\grafvshift by 0.5ex
  \raisebox{\grafvshift}{\includegraphics[height=\grafheight]{images/#2}\xspace}
}

\usepackage{latexsym}
\usepackage{amssymb}
\usepackage{amsmath} 
\usepackage{stmaryrd}
\usepackage{mathtools}
\usepackage{stackrel}

\usepackage{amsthm}
\newtheorem{theorem}{Theorem}[section]
\newtheorem{proposition}[theorem]{Proposition}

\newtheorem{corollary}[theorem]{Corollary}
\theoremstyle{definition}\newtheorem{example}[theorem]{Example}
\theoremstyle{definition}
\theoremstyle{definition}\newtheorem{definition}[theorem]{Definition}
\theoremstyle{definition}
\theoremstyle{definition}\newtheorem{remark}[theorem]{Remark}
\theoremstyle{definition}

\newcommand{\ket}[1]{
    \ensuremath{\left|  #1 \right\rangle}\xspace}

\newcommand{\CX}{\ensuremath{\textsc{cx}}\xspace}

\newcommand{\CNOT}{\CX}

\newcommand{\Rz}{\ensuremath{R_Z}\xspace}
\newcommand{\Rx}{\ensuremath{R_X}\xspace}

\ifx\numericids\undefined

\else

\fi

\newcommand{\zxcalculus}{\textsc{zx}-calculus\xspace}

\usepackage{dsfont}

\newcommand{\tket}{\ensuremath{\mathsf{t}|\mathsf{ket}\rangle}\xspace}

\usepackage{subcaption}
\usepackage{csvsimple}
\usepackage{multirow}
\usepackage{makecell}
\usepackage{algpseudocode}

\usepackage{hyperref}
\usepackage{tikzit}  

\newcommand{\inlinetikzfig}[2][1.0]{
  \dimendef\grafheight=255\dimendef\grafvshift=254
  \setbox0 = \hbox{\scalebox{#1}{\input{#2.tikz}}}
  \grafheight=\the\ht0
  \grafvshift=-0.5\grafheight
  \advance\grafvshift by 0.5ex
  \raisebox{\grafvshift}{\input{#2.tikz}}
}

\newcommand{\inltf}[1]{\InputIfFileExists{#1.tikz}{}{\input{./figures/#1.tikz}}}

\pgfdeclarelayer{background} 
\pgfdeclarelayer{foreground}
\pgfdeclarelayer{edgelayer}
\pgfdeclarelayer{nodelayer}
\pgfsetlayers{background,edgelayer,nodelayer,main,foreground} 

\tikzstyle{halfsize}=[x=0.5cm, y=0.5cm]
\tikzstyle{normalsize}=[]
\tikzstyle{doublesize}=[]

\tikzstyle{(null)}=[]
\tikzstyle{plain}=[]

\usepackage{tikzzx}
\usepackage{tikzcircuits}
\usepackage[undirected]{tikzquanto} 
\usetikzlibrary{calc}

\tikzset{every picture/.style={line width=0.75pt}} 
\tikzset{directed edge/.style={postaction={decorate,decoration={markings,mark=at position 0.5 with {\arrow{>}}}}}}

\usepackage[suppress]{rwd-drafting}
\usepackage{tikz}

\usepackage{afterpage}

\title{A Generic Compilation Strategy for the Unitary Coupled Cluster Ansatz}
\author{
\begin{tabular}{ccccc}
Alexander Cowtan$^{1,}$\thanks{alexander.cowtan@cambridgequantum.com}
& & Will Simmons$^1$ & & Ross Duncan$^{1,2,3}$ \\\\ 
\multicolumn{5}{c}{\footnotesize\makecell{$^1$ \textit{Cambridge Quantum Computing Ltd} \\ \textit{9a Bridge Street, Cambridge, United Kingdom}}} \\
\multicolumn{5}{c}{\footnotesize\makecell{${}^2$ \textit{Department of Computer and Information Sciences}\\
\textit{University of Strathclyde}\\
\textit{26 Richmond Street, Glasgow, United Kingdom}}} \\
\multicolumn{5}{c}{\footnotesize\makecell{${}^3$ \textit{Department of Physics and Astronomy}\\
\textit{University College London}\\
\textit{Gower Street, London WC1E 6BT, United Kingdom}}} \\
\end{tabular}
}
\def\titlerunning{A Generic Compilation Strategy for the Unitary Coupled Cluster Ansatz}
\def\authorrunning{A.~Cowtan, W.~Simmons, and R.~Duncan}

\begin{document}
\maketitle
\begin{abstract}
We describe a compilation strategy for Variational Quantum Eigensolver
(VQE) algorithms which use the Unitary Coupled Cluster (UCC) ansatz,
designed to reduce circuit depth and gate count. This is achieved by
partitioning Pauli exponential terms into mutually commuting sets.
These sets are then diagonalised using Clifford circuits and
synthesised using the phase polynomial formalism. This strategy
reduces \CX depth by 75.4\% on average, and by up to 89.9\%, compared
to naive synthesis for a variety of molecules, qubit encodings and
basis sets. We note that this strategy has potential for other types
of product formula circuits, e.g. for digital Hamiltonian simulation.
\end{abstract}
\section{Introduction}
\label{sec:intro}

Many computational problems in quantum chemistry are classically
intractable for systems which are large and strongly correlated
\cite{Szalay:2012aa}. Instead, quantum algorithms have been proposed
\cite{Kassal:2011aa} to simulate and calculate chemical properties of
such systems. These algorithms leverage useful features of quantum
mechanics to perform calculations which would either take too long or
yield results too inaccurate using the best known classical
algorithms.

However, the resources required by such algorithms tend to be too
large for current quantum computers \cite{Troyer:2015:trotterstepsize}, which are
limited in the number of qubits and the available circuit depth before
decoherence and gate errors overwhelm the system and extracting a
correct result from the noise becomes infeasible. These machines are
known as Noisy Intermediate Scale Quantum (NISQ) devices
\cite{Preskill2018quantumcomputingin}.

A standard approach to reduce the resource requirements enough
to run algorithms successfully on NISQ devices is to only run the
quantum circuit as a subroutine in a larger, classical algorithm
\cite{1367-2630-18-2-023023}. In this model, the quantum circuit
prepares a parameterised state and measures the expectation value of a
relevant operator. The classical routine then performs an optimisation
algorithm, using the expectation value as an objective function, and
attempts to minimise this value with respect to the circuit's
parameters.

The Variational Quantum Eigensolver (VQE) is an archetypal hybrid
quantum-classical algorithm, designed for the estimation of ground
state energies of quantum systems on NISQ devices
\cite{PhysRevA.92.042303}. The expectation value of a molecular
Hamiltonian is the objective function, and VQE employs the variational
principle to approximate the ground state of this Hamiltonian using
the parameterised quantum circuit as an ansatz.

In this paper, we focus on the Unitary Coupled Cluster (UCC) ansatz
\cite{Romero_2018}, which is motivated by the orbital transitions
allowed by the simulated system. We present a compilation strategy for
reducing the major source of error affecting the algorithm: noise of
the quantum device. This compilation strategy increases circuit
fidelity by reducing circuit depth, hence minimising the number of
noisy gates and the qubits' exposure to decoherence.

For NISQ devices, two-qubit gates typically have error rates around an
order of magnitude higher than one-qubit gates, as well as taking 2-5x
as long \cite{Wright:2019aa, Arute:2019aa}. Defining the two-qubit
gate depth as the number of two-qubit parallel layers required to
complete the circuit, we aim to minimise this metric specifically with
our compilation strategy, along with two-qubit gate count. We
approximate the hardware-native two-qubit gate metrics with the
corresponding \CX metrics, assuming that each two-qubit gate must be a
\CX, noting that in certain scenarios this overstates the number of
required gates. Two-qubit gates which are not maximally entangling,
particularly tunable ones, can reduce the number of gates required for
certain algorithms compared to using \CX gates \cite{arute2020quantum,
Nam2019groundstate}.

We begin by partitioning the terms in the UCC ansatz into mutually
commuting sets. We describe a well-known equivalence between this
sequencing problem and graph colouring. We then show that approximate
solutions to this problem enable large-scale synthesis of Pauli
exponentials into one- and two-qubit gates, and propose heuristics for
performing this synthesis to generate low depth circuits.

Our compilation strategy is valid for any ansatz which is generated by
Trotterization of an operator made up of a sum of Pauli tensor
products: this means that it is valid for $k$-UpCCGSD and other
variations on the UCC ansatz, such as UCCGSD \cite{Lee:2019aa} and the
parity-disregarding particle-exchange UCC \cite{xia2020coupled}. It is
also valid for fault-tolerant product formula algorithms for
Hamiltonian simulation \cite{Berry_2006}, although the benefits in
this setting are less clear. Our strategy is not intended for the
hardware efficient ansatz \cite{Kandala:2017aa} or iterative qubit
coupled-cluster ans{\"a}tze \cite{Ryabinkin:2020aa}. The strategy is
generic, and requires no prior knowledge of the qubit encoding, target
operator or basis set beyond the validity condition above. 

We implemented the strategy in \tket \cite{TKETPAPERHERE},
our retargetable compiler, and present benchmarks for a variety of UCC
circuits for realistic molecules to demonstrate empirically that the
strategy significantly reduces the \CX gate count and depth compared
to previous strategies.

\paragraph{Related work:}A similar strategy for optimizing Hamiltonian
simulation circuits was recently presented by van den Berg \& Temme
\cite{Berg:2020aa}. This strategy uses different diagonalisation
methods. In addition, the strategy is intended for fault-tolerant
circuits for Hamiltonian simulation, and the two-qubit gate reduction
is obtained by targeting an ancilla qubit with every \CX from each
diagonal set, as previously described in Hastings et al.
\cite{Troyer:2014:improvedchemistry}, which is impractical for some
NISQ devices. However, a thorough comparison of strategies for Pauli
partitioning and diagonalisation is presented, which can be applied in
the NISQ setting.

\paragraph{Notation:}In order to reason about and represent the
synthesis of Pauli exponentials, we use notation inspired by the
\zxcalculus \cite{Coecke:2009aa}, although our strategy can be
followed without requiring any knowledge of the inference rules of the
calculus. A brief introduction to the \zxcalculus is found in Fagan \&
Duncan \cite{EPTCS287.5}; for a complete treatment see Coecke \&
Kissinger \cite{Coecke2017Picturing-Quant}.

\paragraph{Terminology:}We refer to an $n$-qubit operator of the form
$\{I,X,Y,Z\}^{\otimes n}$ as a \emph{Pauli string}, composed of
\emph{letters} from the alphabet $\{I,X,Y,Z\}$. The \textit{weight} of
a Pauli string is the number of non-$I$ letters.

\section{The Unitary Coupled Cluster Ansatz}
\label{sec:ucc_ansatz}
The UCC ansatz is defined by the excitation of some reference state by
an operator parameterised with coupled cluster amplitudes $\vec{t}$:
\begin{equation}
\ket{\Psi (\vec{t})} = U(\vec{t})\ket{\Phi_0} = e^{T(\vec{t})-T^{\dagger}(\vec{t})}\ket{\Phi_0}
\end{equation}
where operator $T$ is a linear combination of fermionic excitation operators
$\vec{\tau}$ such that the parameterised operator can be rewritten:
\begin{equation}\label{eq:excite-sum}
U(\vec{t}) = e^{\sum_j t_j (\tau_j - \tau^{\dagger}_{j})}
\end{equation}

This parameterised operator cannot be directly implemented on
a gate-based quantum computer. It must be mapped to qubits and
decomposed into native gates. 

In order to generate a quantum circuit, we employ Trotterization,
justified by Lloyd \cite{Lloyd:1996aa}. Here we show the first order
Lie-Trotter expansion:
\begin{equation}\label{eq:trotter-expr}
U(\vec{t}) \approx U_{Trott}(\vec{t}) = (\prod_{j}e^{\frac{t_j}{\rho}(\tau_j-\tau_j^{\dagger})})^{\rho}
\end{equation}
where $\rho$ is the number of Trotter steps. Since our focus is on the
NISQ setting, we will assume that only one Trotter step is taken. It
is straightforward to extend the presented techniques to arbitrary
step size.

To implement the Trotterized expression shown in
Equation~\ref{eq:trotter-expr} on a quantum computer, we map the
$\tau_j$ in the product to operations acting on qubits. This can be
performed using a variety of qubit encodings, such as Bravyi-Kitaev
(BK), Jordan-Wigner (JW) and parity (P) \cite{Steudtner_2018}. These
encodings have different resource requirements and the qubits
represent different physical properties, but regardless of our choice
we obtain:
\begin{equation}\label{eq:excite-expand}
(\tau_j - \tau_j^{\dagger}) = ia_j\sum_k  P_{jk}
\end{equation}
where $a_{j} \in \mathbb{R}$ and $P_{jk} \in \{I,X,Y,Z\}^{\otimes n}$.

It can be shown that the Pauli strings $P_{jk}$ from a given
excitation operator $\tau_j$ always commute under multiplication
\cite{Romero_2018}. This gives a simpler expression for the
Trotterized operator,
\begin{equation}\label{eq:prod-formula}
U_{Trott}(\vec{t}) = \prod_j \prod_k e^{it_ja_jP_{jk}}
\end{equation}
where $e^{it_j a_{j}P_{jk}}$ terms are parameterised with some angle
$t_j$ which will be adjusted by the variational algorithm. We refer to
these terms as \emph{Pauli exponentials}, and relabel our coefficients
$t^{'}_j = t_j a_j$.

Pauli exponentials can be implemented on a quantum computer by
decomposition into one- and two-qubit native gates, discussed in
Section~\ref{sec:reppauli}. These gates are appended to a short,
constant depth circuit generating the reference state. 

\section{Term Sequencing by Graph Colouring}
\label{sec:termseq}

Looking again at Equation~\ref{eq:excite-sum}, note that we can expand
the fermionic excitation operators at this stage into Pauli strings
using Equation \ref{eq:excite-expand}, i.e. our chosen qubit encoding:
\begin{equation}\label{eq:sum-expand}
U(\vec{t}) = e^{i\sum_j  \sum_k t^{'}_j P_{jk}}
\end{equation}
Since addition is commutative we can freely choose the ordering of
$P_{jk}$ terms in this expression.  After
Trotterization the Pauli exponentials do not commute, so it is
sensible at this stage to sequence the Pauli strings in a
beneficial order, such that our Trotterization incurs minimal Trotter
error and our circuit has low resource requirements. The implications
of the ordering of terms for chemical accuracy have been studied by H.
R. Grimsley et al. \cite{Grimsley_2019}. We justify in
Appendix~\ref{app:trot-error} that reducing Trotter error should be a
secondary concern for near-term VQE, and focus on minimising \CX gate
count and depth.

Our strategy to reduce \CX gate count and depth relies on partitioning
the set of Pauli exponentials into a small number of subsets, such
that within a given subset every Pauli exponential commutes.
\footnote{This problem is common in the literature on measurement
reduction \cite{jena2019pauli, crawford2019efficient,
zhao2019measurement, verteletskyi2019measurement}.} This partitioning
problem can be represented as the well-known graph colouring problem.


\begin{figure}
\centering
\input{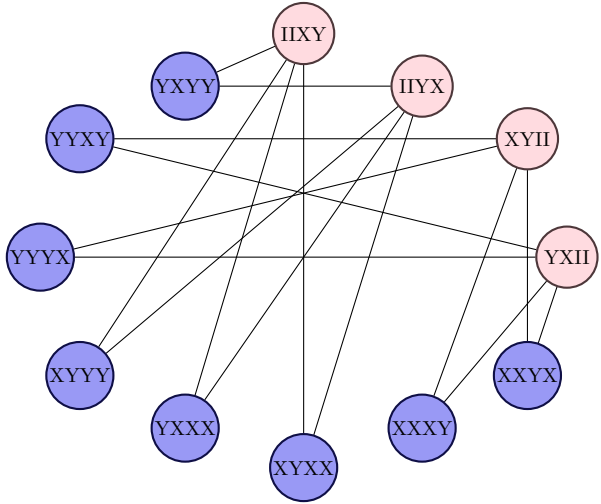}
\caption{Graph colouring to partition Pauli terms into sets of
 mutually commuting strings. While the parameters are not shown, they
 must be tracked for synthesis later.}
\label{fig:pauli_graph1}
\end{figure}

We represent each Pauli string as a vertex in an undirected graph. An
edge is given between any two vertices that correspond to Pauli
strings which anti-commute. Figure~\ref{fig:pauli_graph1} shows an
example of this graph representation.

Finding the minimum number of mutually commuting sets which cover all
vertices in this graph is then equivalent to the \textit{colouring
problem}, a well known NP-hard problem \cite{Garey:1974aa}. In this
instance, the colour assigned to the vertex corresponds to the subset
the corresponding Pauli exponential is placed in, and since no two
adjacent vertices can have the same colour, all Pauli exponentials
within a subset will mutually commute.

We use a simple greedy colouring algorithm to partition the Pauli
strings. The complexity of this algorithm is $\mathcal{O}(m)$, with
$m$ the number of Pauli strings, although building the
anti-commutation Pauli graph in the first place scales as
$\mathcal{O}(m^2n)$, with $n$ the number of qubits.

Once the vertices have been assigned colours, the UCC reference state
$\ket{\Phi_0}$ is prepared and the corresponding Pauli exponential
terms are appended, colour by colour, in lexicographical order. For
example, given the graph colouring solution from
Figure~\ref{fig:pauli_graph1}, a valid ordering of strings is: IIXY,
IIYX, XYII, YXII, XXXY, XXYX, XYXX, XYYY, YXXX, YXYY, YYXY, YYYX.
Neither the order of the sets nor the order of terms within each set
is considered important for optimisation; lexicographical order was an
arbitrary choice.

\section{Pauli Exponentials}
\label{sec:reppauli}

\begin{figure}[t]
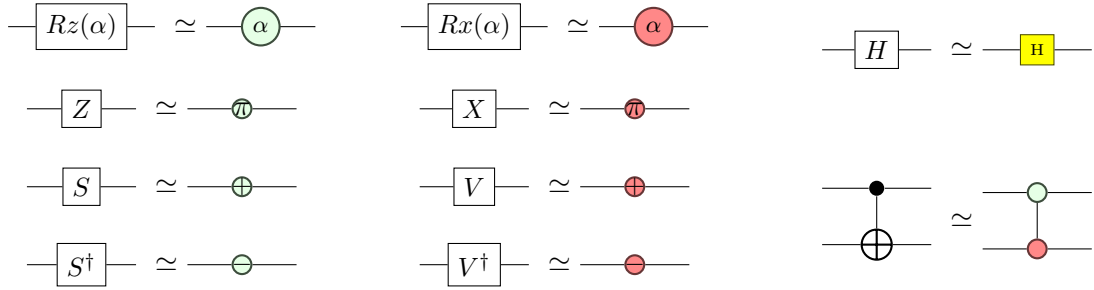

\centering
\begin{tabular}{ccc}
\begin{minipage}{.3\textwidth}
\begin{equation*}
\inltf{GatesRZ} \simeq \inltf{ZXCalcRZ}
\end{equation*}
\end{minipage} &
\begin{minipage}{.3\textwidth}
\begin{equation*}
\inltf{GatesRX} \simeq \inltf{ZXCalcRX}
\end{equation*}
\end{minipage} &
\multirow{2}{*}{
\begin{minipage}{.3\textwidth}
\begin{equation*}
\inltf{GatesH} \simeq \inltf{ZXCalcH}
\end{equation*}
\end{minipage}
}
\\
\begin{minipage}{.25\textwidth}
\begin{equation*}
\inltf{GatesZ} \simeq \inltf{ZXCalcZ}
\end{equation*}
\end{minipage} &
\begin{minipage}{.25\textwidth}
\begin{equation*}
\inltf{GatesX} \simeq \inltf{ZXCalcX}
\end{equation*}
\end{minipage} & \\
\begin{minipage}{.25\textwidth}
\begin{equation*}
\inltf{GatesS} \simeq \inltf{ZXCalcS}
\end{equation*}
\end{minipage} &
\begin{minipage}{.25\textwidth}
\begin{equation*}
\inltf{GatesV} \simeq \inltf{ZXCalcV}
\end{equation*}
\end{minipage} &
\multirow{2}{*}{
\begin{minipage}{.25\textwidth}
\begin{equation*}
\inltf{GatesCX} \simeq \inltf{ZXCalcCX}
\end{equation*}
\end{minipage}
}
\\
\begin{minipage}{.25\textwidth}
\begin{equation*}
\inltf{GatesSdg} \simeq \inltf{ZXCalcSdg}
\end{equation*}
\end{minipage} &
\begin{minipage}{.25\textwidth}
\begin{equation*}
\inltf{GatesVdg} \simeq \inltf{ZXCalcVdg}
\end{equation*}
\end{minipage} &
\end{tabular}
\caption{Common circuit gates and their representations in the
scalar-free \zxcalculus. The $S$ gate corresponds to
$\Rz(\frac{\pi}{2})$, and the $V$ gate to
$\Rx(\frac{\pi}{2})$. }
\label{fig:zx_gates}
\end{figure}

A translation of relevant gates between the quantum circuit
and \zxcalculus formalisms is given in Figure~\ref{fig:zx_gates}. 

Recall the notation of \textit{phase gadgets} $\Phi_n(\alpha)$,
equivalent to the operator $e^{-i\frac{\alpha}{2} Z^{\otimes n}}$.
These gadgets were described in Kissinger \& van de Wetering
\cite{Kissinger:2019aa}, and have a natural representation in the
\zxcalculus.

\begin{definition}\label{def:phasegadget}
  In \zxcalculus notation we have:
  \[
  \Phi_n(\alpha) := \inltf{PhaseGadgetDef} = \inltf{PhaseGadgetDecomp}
  \]
\end{definition}

The algebra for phase gadgets and alternate decompositions into one-
and two-qubit gates are given in Appendix~\ref{app:phase-gadgets}.
Note that $\Phi_1(\alpha) = \Rz(\alpha)$.

The correspondence between phase gadgets and Pauli-$Z$ exponentials
generalises to any Pauli exponential $e^{-i\frac{\alpha}{2} P}$, by
conjugating the phase gadget with single-qubit Clifford gates. We
recall the \textit{Pauli gadget} diagrammatic notation for the Pauli
exponential from Cowtan et al. \cite{Cowtan:2019aa}.

\begin{definition} Pauli exponentials are represented succinctly as:
    \[
    e^{-i\frac{\alpha}{2} I X Y Z} = {\input{PhaseGadgetIXYZ.tikz}} = {\input{PauliExpDef.tikz}}
    \]
\label{def:pauli_exp_gadget}
\end{definition}

The red, mixed-colour, and green boxes respectively represent the
Pauli gadget acting on a qubit in the $X$, $Y$, and $Z$ bases. These
are formed by a phase gadget on the qubits (generating $Z$-only
interactions), then conjugating the qubits with appropriate
single-qubit Cliffords.

Clifford gates may be commuted through Pauli gadgets, but may incur a
phase flip or basis change. The exhaustive set of diagrammatic rules
required to perform this procedure for relevant Clifford gates are
shown in Appendix~\ref{app:cliff-commutation}, although they are
simple to calculate using linear algebra.

The definitions above imply a naive method of circuit synthesis for
Pauli gadgets.  For a set of $m$ Pauli gadgets over $n$ qubits, this
naive synthesis requires $\mathcal{O}(nm)$ \CX gates. More, precisely,
we require at most $2m(n-1)$, if all the Pauli strings are maximal
weight.  This gives the baseline performance against which we will
compare the method introduced in the next section.


\section{Set-based Synthesis}
\label{sec:setsynth}

The effect of the transformations in Section~\ref{sec:termseq} is to
partition our ansatz into large commuting sets of Pauli gadgets.  The
next step is to synthesise circuits for these sets, while minimising
the \CX overhead.

The approach we propose here has two steps:
\begin{enumerate}
\item Diagonalisation: every Pauli gadget in a given commuting set is
simultaneously converted into a phase gadget by conjugating with an
appropriate Clifford circuit. 
\item Phase gadget synthesis: the resulting phase gadgets are
converted into \CX and $\Rz$ gates using the well-studied \textit{phase
polynomial} formalism \cite{Amy2014Polynomial-Time}. 
\end{enumerate}
While diagonalisation incurs a gate overhead, in practice we find that
the gate reduction from synthesising using this technique more than
makes up for this overhead.
Figure~\ref{fig:complexities} summarises the relevant complexities of
the different subroutines in our strategy. 

\begin{figure}[t]
\centering
\begin{tabular}{l|c|c}
      & Time complexity & \CX complexity  \\ \hline
      Graph Colouring  & $\mathcal{O}(m^2n)$  & - \\
      Diagonalisation     & $\mathcal{O}(mn^3)$  & $\mathcal{O}(n^2)$ \\
      GraySynth \cite{Amy_2018, AmyEmail:2020aa} & $\mathcal{O}(mn^3)
      $  & $\mathcal{O}(mn)$ \\ \hline
\end{tabular}
\caption{Summary of subroutine complexities, where $m$ is the total
  number of Pauli exponentials and $n$ is the number of qubits.  Time
  complexity refers to the compilation time, while \CX complexity is
  defined as the maximum number of \CX gates required for circuit
  synthesis. Graph colouring does not perform circuit synthesis so has no \CX complexity.}
\label{fig:complexities}
\figureline
\end{figure}

\subsection{Diagonalisation}
\label{sec:diag}

Phase gadgets -- that is, Pauli gadgets whose Pauli strings contain
only the letters $Z$ and $I$ -- define unitary maps which are diagonal
in the computational basis.  For this reason, we'll call a \emph{set}
of Pauli gadgets diagonal when it contains only phase gadgets. Abusing
terminology slightly, given a set of Pauli gadgets, we call a
\emph{qubit} diagonal over that set when the Pauli letter for that
qubit is either $Z$ or $I$ for every Pauli string in the set.
Evidently, if every qubit is diagonal then the set as a whole is too.

A set $S$ of commuting Pauli gadgets can be simultaneously
diagonalised using only Clifford operations.  Our goal is to find
a Clifford circuit $C$ and a diagonal set $S'$ such that 
\begin{equation}\label{eq:basic-diag-relation}
S = CS'C^{\dagger}
\end{equation}
where $C$ is as small as possible.  Several methods have been proposed
\cite{jena2019pauli,crawford2019efficient,Scott-Aaronson:2004yf,
  Maslov2017Shorter-stabili} to compute a suitable polynomially-sized
circuit $C$.  

Note that, since $[A,B] = 0 \iff [UAU^{\dagger},UBU^{\dagger}] = 0$
for unitaries $A$, $B$ and $U$, conjugating the gadgets preserves
commutation, so the required $C$ can be constructed by conjugation
with Cliffords.  Below, we use this approach on \emph{compatible
  pairs} of qubits, where one can qubit can be used to diagonalise the
other.  In the worst case $C$ has $\mathcal{O}(n^2)$ \CNOT gates;
however in practice, on realistic examples of UCC ansatz circuits, we
find our method typically produces Clifford diagonalisers much smaller
than the asymptotic worst case.

\begin{remark}\label{rem:diag-complexity-related-works}
Jena et al.~\cite{jena2019pauli} presented an algorithm guaranteed to
give such a $C$ for qudits of any dimension, of size quadratic in
the number of qudits.
For $m$ Pauli gadgets, Crawford et al.~\cite{crawford2019efficient}.
recently presented two efficient constructions of $C$ with a bound of
$mn-m(m+1)/2$ and $\mathcal{O}(mn/\log m)$ \CX gates respectively,
when $m < n$.
When $m \geq n$, the construction
provided by Aaronson \& Gottesman requires $\mathcal{O}(n^2/\log n)$
\CX gates \cite{Scott-Aaronson:2004yf, Maslov2017Shorter-stabili}.
\end{remark}

\subsubsection{Diagonalising a compatible pair}

\begin{figure}
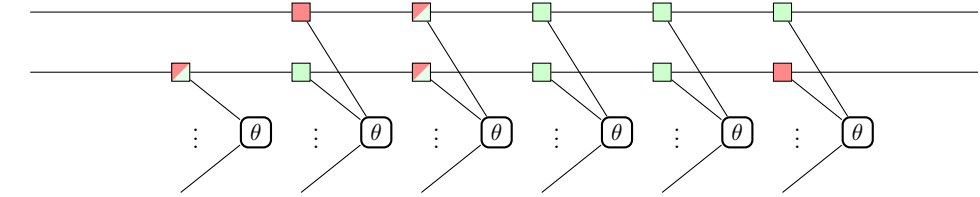
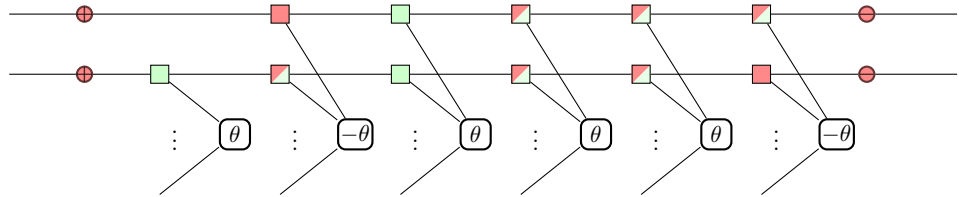
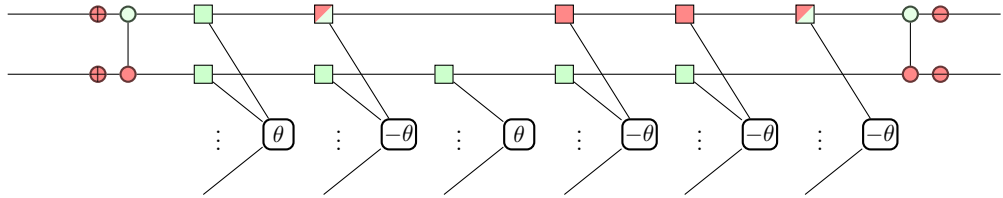
\label{fig:theorem_example}
    \begin{subfigure}[b]{\textwidth}
    \begin{equation*}
    \inltf{theorem_example1}
    \end{equation*}
    \caption{A compatible pair.}
    \label{fig:theorem_example1}
    \end{subfigure}
    \vspace{6mm}

    \begin{subfigure}[b]{\textwidth}
    \begin{equation*}
    \inltf{theorem_example2}
    \end{equation*}
    \caption{Conjugation with appropriate single-qubit Cliffords.}
    \label{fig:theorem_example2}
    \end{subfigure}
    \vspace{6mm}

    \begin{subfigure}[b]{\textwidth}
    \begin{equation*}
    \inltf{theorem_example3}
    \end{equation*}
    \caption{Conjugation with \CX gates to diagonalise the second qubit.}
    \label{fig:theorem_example3}
    \end{subfigure}
    \vspace{6mm}

    \caption{Application of Theorem~\ref{thm:pauli-chain} to diagonalise a qubit.}
\end{figure}


In the following, let $S$ be a set of $m$ commuting Pauli gadgets
acting on $n$ qubits, and let $\sigma_{kl}$ denote the Pauli letter on
qubit $k$ from gadget $l$.

\begin{definition}\label{def:compatible-pair}
  Let $i,j \in \{1, \ldots, n\}$ with $i\neq j$.  Qubits $i$ and $j$
  are called \emph{compatible} in $S$ if the following relation holds:
  \begin{equation}\label{eq:pauli-diag-condition}
    \exists A, B \in \{X,Y,Z\} \:\: s.t. \:\: \forall l \in
    \{1,...,m\}, \sigma_{il} \in \{I,A\} \iff \sigma_{jl} \in \{I,B\} \;\;;
  \end{equation}
  In this case $i$ and $j$ are called a \emph{compatible pair}.
\end{definition}

\begin{theorem}\label{thm:pauli-chain}
  If qubits $i$ and $j$ form a compatible pair in $S$ then one of them can be
  can be diagonalised by conjugating $S$ with a single \CNOT and at most two
  single qubit Cliffords acting on qubits $i$ and $j$.
  \begin{proof}
    Without loss of generality, assume $i=1$ and $j=2$, and let $P_2$
    be a 2-qubit Pauli string.  Applying the transformation $ P_2
    \mapsto \CNOT \circ P_2 \circ \CNOT $ (with the second qubit the
    target) will diagonalise the second qubit when $P_2 \in \{II, IZ,
    XX, XY, YX, YY, ZI, ZZ\}$.  This set satisfies the property
    $\sigma_{1} \in \{Z,I\} \iff \sigma_{2} \in \{Z,I\}$, and all
    other sets of 2-qubit Paulis which satisfy this property are
    subsets of this one. (Note that the control qubit will be diagonal
    after conjugation iff it was diagonal before.)  Conjugating the
    first and/or second qubit by an additional single qubit Clifford
    allows this relation to be generalised to any pair of Paulis as in
    (\ref{eq:pauli-diag-condition}), giving the result.
  \end{proof}
\end{theorem}


If $i$ and $j$ are a compatible pair, then the specific values of $A$
and $B$ in relation (\ref{eq:pauli-diag-condition}) determine which
single-qubit Cliffords are required before conjugation by \CX gates
will diagonalise a qubit.  An example is shown in
Figure~\ref{fig:theorem_example1}.  The first two qubits are
compatible, with $A = Y$ and $B = Y$, which implies that $V$ and
$V^{\dagger}$ gates are required to prepare the second qubit for
diagonalisation as shown in Figure~\ref{fig:theorem_example2}.  The
diagonalisation is completed by \CNOT conjugation as shown in
Figure~\ref{fig:theorem_example3}. 

Applying Theorem~\ref{thm:pauli-chain} to compatible pairs of qubits
is the key subroutine in our diagonalisation algorithm, described in
the next section.

\begin{corollary}
    For any commuting set of $m$ gadgets over $n$ qubits, if $m < 4$ a
    Clifford circuit $C$ exists which diagonalises this set of gadgets
    using at most $n-1$ \CX gates.
    \begin{proof}
    See Appendix~\ref{app:corollary54proof}.
    \end{proof}
\end{corollary}
\begin{corollary}
    For any commuting set of gadgets over $n$ qubits, if $n < 5$ a
    Clifford circuit $C$ exists which diagonalises this set of gadgets
    using at most $n-1$ \CX gates.
    \begin{proof}
    See Appendix~\ref{app:corollary55proof}.
    \end{proof}
\end{corollary}

\subsubsection{Diagonalising a commuting set}

\begin{figure}
\begin{algorithmic}

\Function{GadgetDiag}{$S$}
    \State $Q \gets $ Qubits($S$) 
    \State $C \gets $ EmptyCircuit(Q)
    \While {$Q$ non-empty} 
    \State ($S$, $Q$, $C$) $\gets $ UpdateSingleQubits($S$, $Q$, $C$)
    \If {$Q$ empty}
    \State \textbf{break}
    \EndIf
    \State $Q^{'} \gets Q$
    \State ($S$, $Q$, $C$) $\gets $ UpdatePairQubits($S$, $Q$, $C$)
    \If {$Q = Q^{'}$}
    \State ($S$, $Q$, $C$) $\gets $ GreedyDiagonalisation($S$, $Q$, $C$)
    \EndIf
    \EndWhile
    \State \textbf{return} ($S$, $C$)
\EndFunction

\Function{UpdateSingleQubits}{$S$, $Q$, $C$}
\For {$q \in Q$}
\State $p$ $\gets $ FindCommonPauli($S$, $q$) \Comment{$p$ : Maybe Pauli}
\If {$p \neq $ None}
\State $S \gets $ UpdateGadgetsSingleQubit($S$, $p$, $q$)
\State $Q \gets Q \setminus \{ q \} $
\State $C \gets $ AddCliffordsSingleQubit($C$, $p$, $q$)
\EndIf
\EndFor
\State \textbf{return} ($S$, $Q$, $C$)
\EndFunction
\vspace{3mm}

\Function{UpdatePairQubits}{$S$, $Q$, $C$}
\For {$q_a \in Q$}
\For {$q_b \in Q \setminus \{q_a\}$}
\State ($p_a$, $p_b$) $\gets $ FindValidPaulis($S$, $q_a$, $q_b$)
\Comment{($p_a$, $p_b$) : Maybe Pair Pauli}
\If {($p_a$, $p_b$) $\neq$ None}
\State $S \gets $ UpdateGadgetsPairQubit($S$, $p_a$, $p_b$, $q_a$, $q_b$)
\State $Q \gets Q \setminus \{ q_b \} $
\State $C \gets $ AddCliffordsPairQubit($C$, $p_a$, $p_b$, $q_a$, $q_b$)
\State \textbf{return} ($S$, $Q$, $C$)
\EndIf
\EndFor
\EndFor
\State \textbf{return} ($S$, $Q$, $C$)
\EndFunction
\vspace{3mm}

  \end{algorithmic}
 \caption{Diagonalisation algorithm}
 \label{alg:diagonalise}
\end{figure}

This section describes our method for diagonalising a set of commuting
Pauli gadgets.  The basic approach is to repeatedly apply three
methods which diagonalise a single qubit : 
\begin{enumerate}
\item Diagonalise the trivially diagonalisable qubits
\item Diagonalise qubits in compatible pairs
\item Synthesise a single gadget to diagonalise one of its qubits.
\end{enumerate}
Detailed pseudo-code for the algorithm\footnote{
  We have omitted the greedy diagonalisation method from the
  pseudo-code, as it is straightforward.
} is presented in Figure~\ref{alg:diagonalise}.  The overall time
complexity for this algorithm is $\mathcal{O}(mn^3)$, with $m$ the
number of Pauli gadgets in the commuting set and $n$ the number of
qubits.

To make the algorithm clearer, we'll go through a worked example in
Figures~\ref{fig:set_exampleA} and \ref{fig:set_exampleB}.  Initially
we have the mutually commuting gadgets shown in
Figure~\ref{fig:set_example1}, corresponding to the Pauli strings
$IXZIZ$, $IYIZY$, $XXIYI$, $YYXII$, $ZIYXX$, $ZXIZZ$, and $ZYZIY$.  We
proceed as follows.
\begin{enumerate}
\item First, check whether there is a trivially diagonalisable qubit:
that is, a qubit $i$ for which $\exists P \in \{ X, Y, Z\}$ s.t.
$\forall l, \sigma_{il} \in \{I, P\} $. Any such qubits may be
diagonalised with only single-qubit Clifford gates, and ignored from
now on. This check takes time $\mathcal{O}(mn)$.
Figure~\ref{fig:set_example1} contains no such qubits.

\item Now search for a compatible pair of qubits
  (Defn.~\ref{def:compatible-pair}) 
Theorem~\ref{thm:pauli-chain} for any choice of Paulis $A$ and $B$,
and apply the conjugation of Theorem \ref{thm:pauli-chain} to
diagonalise a qubit and remove it from consideration.  The choice of
qubit within the pair is arbitrary.
This search can be performed in $\mathcal{O}(mn^2)$.  The example of 
Figure~\ref{fig:set_example1} does not contain a compatible pair.

\item If no compatible pair is found, we adopt a greedy approach as a
backup strategy. In $\mathcal{O}(m)$ time, find the Pauli string with
the lowest weight; if there are multiple pick arbitrarily. Conjugate
the corresponding Pauli gadget with single-qubit Clifford and \CX
gates to convert the Pauli string to $II\ldots IZ$, demonstrated in
Figure~\ref{fig:set_example2}. Then, commute the Clifford gates
through the rest of the gadgets, as shown in
Figure~\ref{fig:set_example3}, until all Clifford gates are outside
the adjacent Pauli gadgets. Every gadget must still commute with the
$II\ldots IZ$ string, and therefore the last qubit must be diagonal.
This is a similar method to Jena et al. \cite{jena2019pauli}.
\end{enumerate}
These steps are repeated until all qubits are diagonal over the set
of Pauli gadgets. Following our example, we find that
Figure~\ref{fig:set_example3} has the same two-qubit chain on the
first and second qubits as our example from
Figure~\ref{fig:theorem_example1}, and can therefore be diagonalised
in the same way, resulting in the circuit in
Figure~\ref{fig:set_example5}. The backup strategy is not required for
the remaining qubits. See Figure~\ref{fig:set_example6} for the
circuit after full diagonalisation.

Since each iteration will diagonalise at least one qubit,
$\mathcal{O}(n)$ repetitions are required,
so Algorithm~\ref{alg:diagonalise} has time complexity
$\mathcal{O}(mn^3)$.   
In the worst case, the greedy approach is required repeatedly, so $C$
will require at most $\frac{1}{2}n(n-1)$ \CX gates. If the greedy
approach is not required at all, $C$ will require at most $n-1$ \CX
gates. For our small, 5-qubit example circuit, the greedy approach was
required at one iteration, and $C$ used 5 \CX gates.

\afterpage{
\begin{figure}[h]
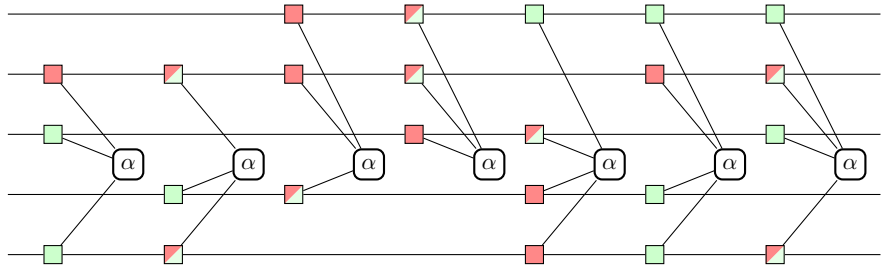
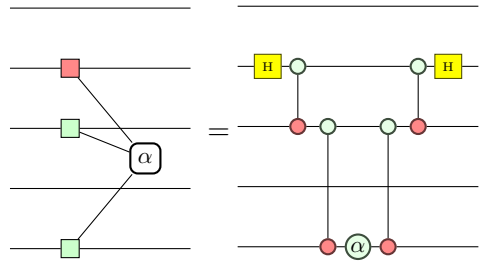
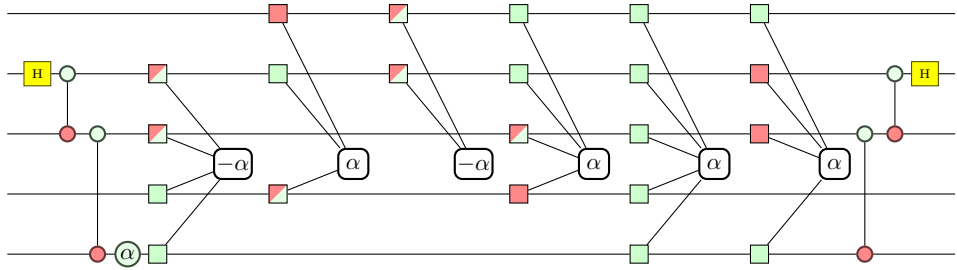
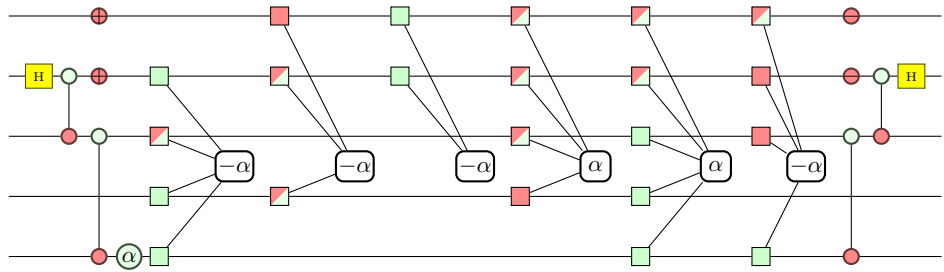

    \begin{subfigure}[b]{\textwidth}
    \begin{equation*}
    \inltf{commuting_set1}
    \end{equation*}
    \caption{Example set of adjacent commuting Pauli gadgets.}
    \label{fig:set_example1}
    \end{subfigure}
    \vspace{6mm}

    \begin{subfigure}[b]{\textwidth}
    \begin{equation*}
    {\inltf{decomp_gadget1}} = {\inltf{decomp_gadget2}}
    \end{equation*}
    \caption{Leftmost Pauli gadget before and after decomposition.}
    \label{fig:set_example2}
    \end{subfigure}
    \vspace{6mm}

    \begin{subfigure}[b]{\textwidth}
    \begin{equation*}
    {\inltf{decomp_set2}}
    \end{equation*}
    \caption{Pauli gadget set after commuting Cliffords through.}
    \label{fig:set_example3}
    \end{subfigure}
    \vspace{6mm}
    
    \begin{subfigure}[b]{\textwidth}
    \begin{equation*}
    {\inltf{decomp_set_YY_1}}
    \end{equation*}
    \caption{Theorem~\ref{thm:pauli-chain} is satisfied by $A=Y$ and $B=Y$ for the first two qubits. Single-qubit rotations are applied to convert to the $Z$-basis.}
    \label{fig:set_example4}
    \end{subfigure}

    \caption{Set diagonalisation example.}
    \label{fig:set_exampleA}
\end{figure}

\begin{figure}[h]
    \centering
    \begin{subfigure}[b]{\textwidth}
    \begin{equation*}
    {\inltf{decomp_set_YY_2}}
    \end{equation*}
    \caption{Diagonalise the second qubit with a \CX.}
    \label{fig:set_example5}
    \end{subfigure}
    \vspace{10mm}

    \begin{subfigure}[b]{\textwidth}
    \centering
    {\inltf{decomp_set_YY_4}}
    \caption{Repeat the procedure for the remaining qubits to fully diagonalise the set.}
    \label{fig:set_example6}
    \end{subfigure}
    \vspace{10mm}

    \begin{subfigure}[b]{\textwidth}
    \begin{equation*}
    {\inltf{decomp_set_YY_5}}
    \end{equation*}
    \caption{Convert phase gadgets to \CX and $\Rz$ gates using GraySynth \cite{Amy_2018}.}
    \label{fig:set_example7}
    \end{subfigure}
    \caption{Set diagonalisation example (cont'd).}
    \label{fig:set_exampleB}
\end{figure}
\clearpage
}

\subsection{Phase Polynomials}
\label{sec:phasepoly}

After diagonalisation, we have a circuit $CS'C^\dag$ where the
interior section $S'$ is comprised entirely of phase gadgets,
with each gadget acting on a different set of qubits. This
phase gadget circuit can be expressed as a \emph{phase polynomial}.

\begin{proposition}[Nam et al. \cite{Nam:2018aa}]
\label{prop:phase-poly}
Let $D$ be a quantum circuit containing only \CX gates and the gates
$\Rz(\theta_1)$, $\Rz(\theta_2)$,..., $\Rz(\theta_m)$. The action of $D$
on a basis state $\ket{x_1,x_2...x_n}$ has the form:
\begin{equation}\label{eq:phase-poly-i}
D\ket{x_1,x_2...x_n} = e^{ip(x_1,x_2,...,x_n)}\ket{h(x_1,x_2,...,x_n)}
\end{equation}
where $h(x_1,x_2,...,x_n)$ is a linear reversible function and
\begin{equation}\label{eq:phase-poly-ii}
p(x_1,x_2,...,x_n) = \sum_{i=1}^{m} \theta_i f_i(x_1,x_2,...,x_n)
\end{equation}
is a linear combination of linear Boolean functions $f_i$: $\{0,1\}^n$
$\rightarrow$ $\{0,1\}$. 
\end{proposition}

\begin{definition}\label{def:phase-poly}
  Given $D$ as above, the \emph{phase polynomial} of circuit $D$ is
  $p(x_1,x_2,...,x_n)$, and each $f_i$ is called a \textit{parity}.
\end{definition}

\begin{example}\label{ex:phase-poly}
The circuit shown below has the required form.
\[
\inltf{phase_poly_example} 
\qquad \equiv \qquad
\ket{q_1,q_2} \ \mapsto \ {e^{i \alpha(q_1 \oplus q_2)} \ket{q_1,q_2}}
\]
We can read that the corresponding phase polynomial is
$p(q_1,q_2) = \alpha(q_1 \oplus q_2)$, defined on the single parity
$q_1 \oplus q_2$.
\end{example}

Phase polynomials have been studied for the purposes of circuit
optimisation\footnote{The representation of phase gadget circuits as
  phase polynomials has also inspired circuit optimisation techniques
  \cite{Beaudrap:2019aa}.  }  
\cite{Nam:2018aa, Maslov2017Shorter-stabili, Amy2014Polynomial-Time,
  6516700}.  Phase polynomial \emph{synthesis} refers to the task of
generating a circuit over a chosen gate set, usually \CNOT and
$\Rz(\theta)$,  which implements a given phase polynomial with minimal
resources.  Optimal synthesis is NP-complete in specific cases
\cite{Amy_2018}, but the time complexity of the general case remains open.  
In practice, heuristic methods such as the GraySynth algorithm of Amy
et al. \cite{Amy_2018} can achieve significant reductions in \CNOT count.

The circuit of Example \ref{ex:phase-poly} can be equivalently written
as a phase gadget over two qubits.  In fact, every $n$-qubit phase
gadget is equivalent to a phase polynomial with a single term in the
summation of Eq.~(\ref{eq:phase-poly-ii}), so each phase gadget
corresponds to a parity $f_i$ and a rotation $\theta_i$.
\[
\inltf{gadget_poly_n}
\qquad \equiv \qquad
\ket{q_1,q_2,...,q_n}
\ \mapsto \ 
{e^{i \alpha \big( \bigoplus_{j=1}^{n}q_j \big)} \ket{q_1,q_2,...,q_n}}
\]
More generally, a circuit comprising only phase gadgets can be
represented by a phase polynomial 
where the linear reversible function $h(x_1,x_2,...,x_n)$ of
Eq.~(\ref{eq:phase-poly-i}) is the identity.  This allows us to use
techniques based on phase polynomials to synthesise a circuit for $S'$.

While any synthesis method could be used, for the results described
here we chose the heuristic GraySynth method \cite{Amy_2018} because
it produces an efficient circuit\footnote{
  Across a suite of Clifford+$T$ benchmark circuits, the
  implementation of Amy et al. reduced the \CX gate count by 23\%
 with a maximum reduction of 43\% \cite{Amy_2018}.
}
at reasonable computational cost.  If a specific qubit architecture
was required, then an architecture-aware synthesis method would be more
appropriate \cite{Nash:2019aa,Arianne-Meijer-van-de-Griend:2020aa}.
GraySynth runs in time $\mathcal{O}(mn^3)$, and requires a maximum of
$\mathcal{O}(mn)$ \CX gates when the linear reversible function $h$ is
identity \cite{AmyEmail:2020aa}.  For reasons of space we omit the
algorithm.






Returning to the running example, the synthesised circuit generated
from the interior phase gadgets is shown in
Figure~\ref{fig:set_example7}. Using a naive decomposition, as
described in Definitions~\ref{def:phasegadget} and
\ref{def:pauli_exp_gadget}, the initial set from
Figure~\ref{fig:set_example1} would have required 34 \CX gates, and 34
\CX depth. Our strategy has reduced the \CX count to 22, and the \CX
depth to 18.

\section{Results and Discussion}
\label{sec:results}

We implemented our strategy in \tket \cite{TKETPAPERHERE},
our retargetable compiler. We benchmarked this implementation on a
suite of ansatz circuits for electronic structure UCCSD (Unitary
Coupled Cluster Singles and Doubles) VQE problems. We included the
molecules $\mathrm{H}_2$, $\mathrm{H}_4$, $\mathrm{H}_8$,
$\mathrm{LiH}$, $\mathrm{BeH}_2$, $\mathrm{NH}$,
$\mathrm{H}_2\mathrm{O}$, $\mathrm{CH_2}$, $\mathrm{NH}_3$,
$\mathrm{HNO}$, $\mathrm{HCl}$, $\mathrm{N}_2$, $\mathrm{C}_2$,
$\mathrm{H}_2\mathrm{CO}$, $\mathrm{CO_2}$ and $\mathrm{C}_2
\mathrm{H}_4$ in the `sto-3g' basis set. For the smaller molecules, we
also used the `631g' basis. We tested using the Bravyi-Kitaev (BK),
Jordan-Wigner (JW) and parity (P) encodings.

The comparisons made are: \footnote{We would additionally like to
compare to the low-rank decomposition methods of Motta et al.
\cite{motta2018low}, as the circuit depths and gate counts are stated
to have lower complexity than the standard method described herein.
However, we could not obtain a working implementation of the method.
We would also like to compare to van den Berg \& Temme
\cite{Berg:2020aa}, but data is available only for Hamiltonian
simulation circuits.}
\begin{enumerate}
\item Naive decomposition: circuits generated from
Equation~\ref{eq:prod-formula} by decomposing Pauli gadgets naively
into \CX and single-qubit gates, as described in
Section~\ref{sec:reppauli}.
\item Pairwise synthesis: circuits generated by graph colouring and
then synthesising Pauli gadgets within a set in a pairwise manner with
\CX balanced trees using the methods from Cowtan et al.
\cite{Cowtan:2019aa}.
\item Set synthesis: our full compilation strategy. Graph colouring,
diagonalisation and phase polynomial synthesis.
\item Templated lexicographical operator sequence (TLOS): for ansatz
circuits generated using the JW encoding we compare against a mock
implementation of the best known previous strategy for JW circuit
synthesis: operator sequencing methods from Hastings et al.
\cite{Troyer:2014:improvedchemistry} allowing for \CX cancellation
between excitations, with templated excitation operators from Nam et
al.~\cite{Nam2019groundstate} for low \CX count excitations
\footnote{We do not allow the use of ancilla qubits for this method,
which Hastings et al. showed can reduce \CX overhead significantly.
Additionally, Nam et al. used a bosonic excitation technique relating
molecular and spin orbitals, which we do not include here.}. 
We are not aware of similar strategies for the BK or P encoding.
\end{enumerate}

Circuits in our test set were chosen to have a \CX count and depth of
less than $10^6$ when decomposed naively. All results were obtained
using \texttt{pytket v0.5.5}, on a machine with a 2.3~GHz Intel Core
i5 processor and 8~GB of 2133~MHz LPDDR3 memory, running MacOS
Mojave~v10.14.

A benchmark script for reproducing the results, along with the input
operators, can be found at
\url{https://github.com/CQCL/tket_benchmarking/tree/master/compilation_strategy}.
The methodology for generating and serialising these operators is
described in Appendix~\ref{app:qboperator}.

\begin{figure}
\centering

\begin{subfigure}[b]{\textwidth}
\begin{tabular}{cc}
\includegraphics[scale=0.5]{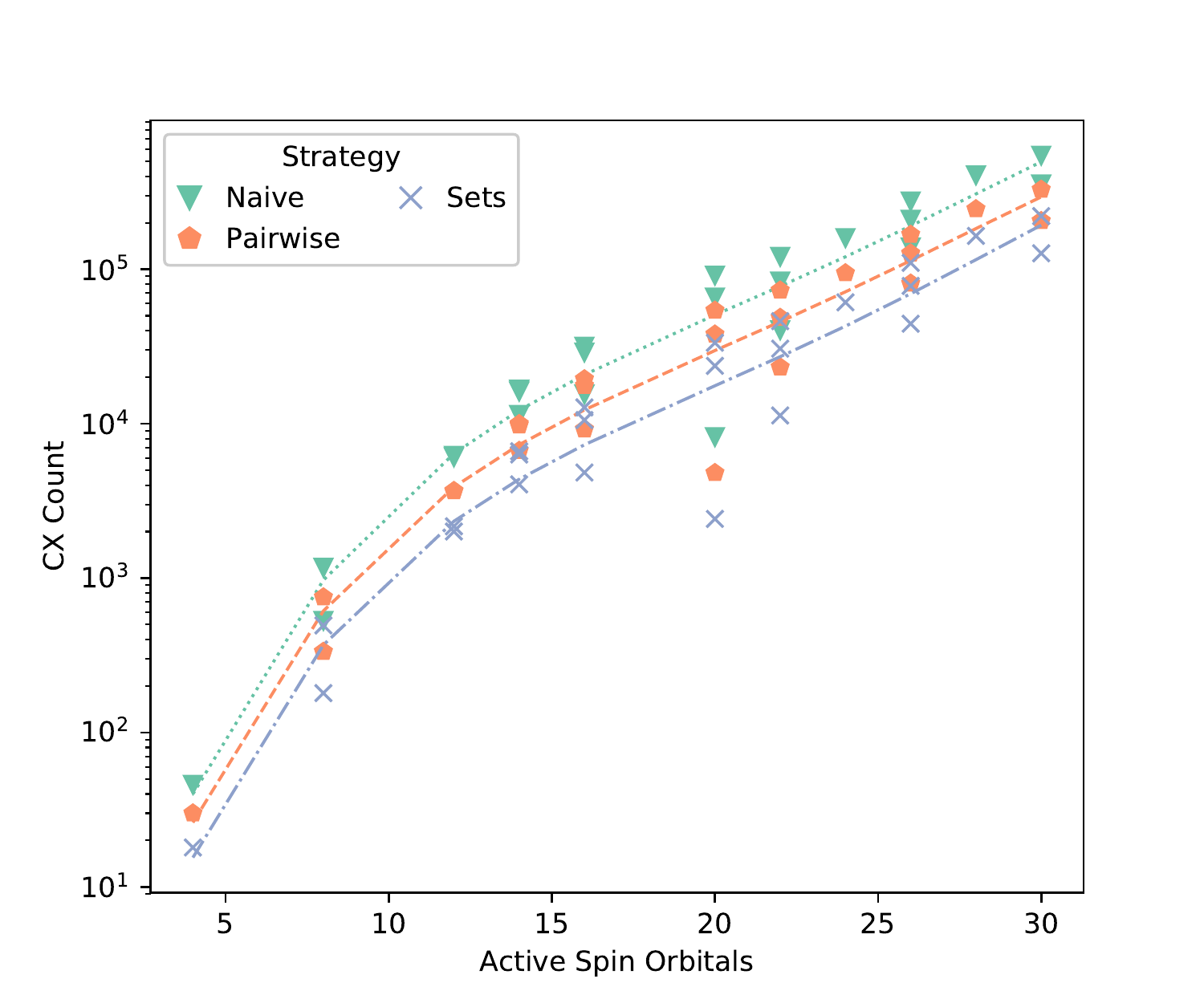} &
\includegraphics[scale=0.5]{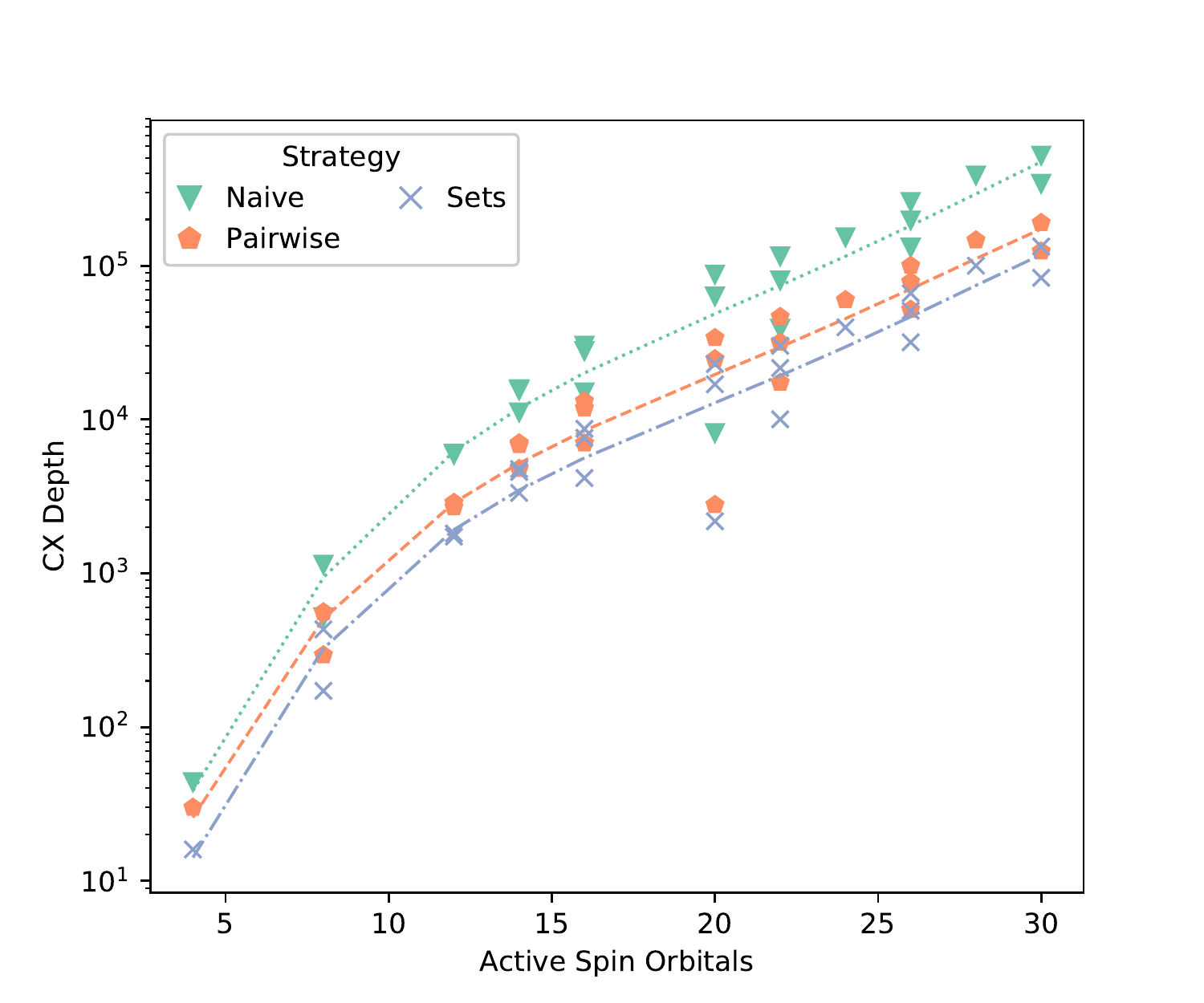}
\end{tabular}
\caption{Bravyi-Kitaev qubit encoding.}
\end{subfigure}

\begin{subfigure}[b]{\textwidth}
\begin{tabular}{cc}
\includegraphics[scale=0.5]{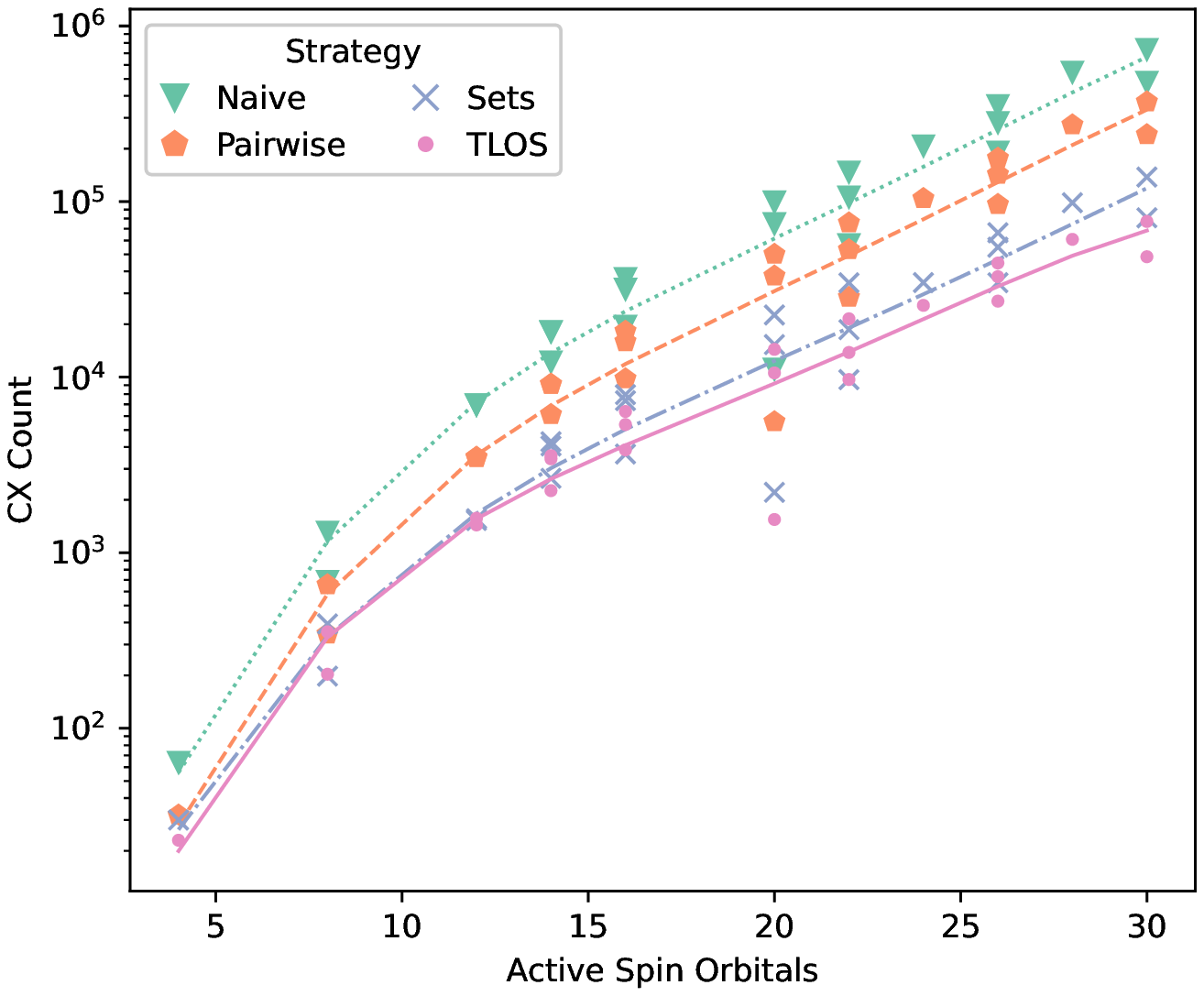} &
\includegraphics[scale=0.5]{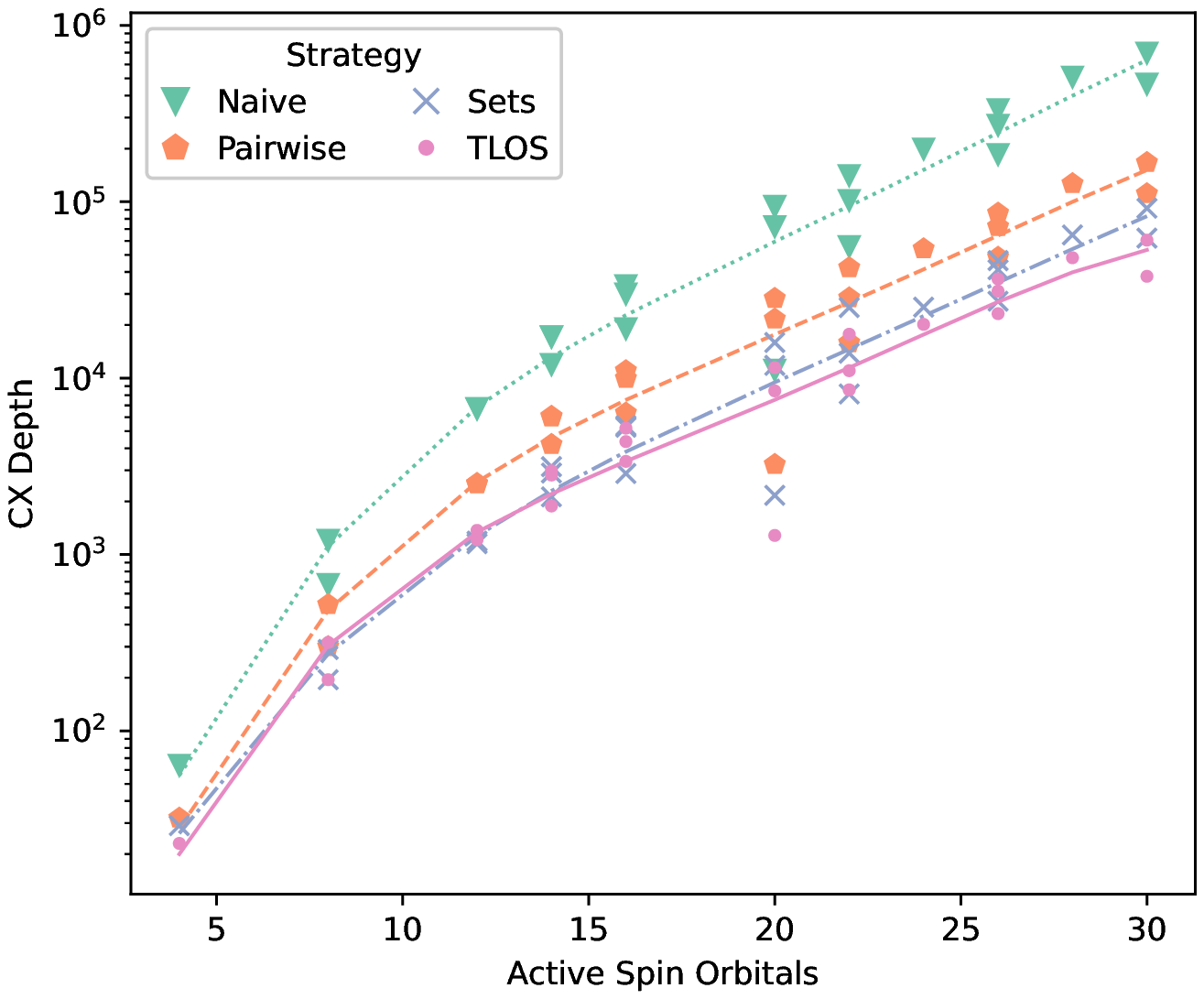}
\end{tabular}
\caption{Jordan-Wigner qubit encoding.}
\end{subfigure}

\begin{subfigure}[b]{\textwidth}
\begin{tabular}{cc}
\includegraphics[scale=0.5]{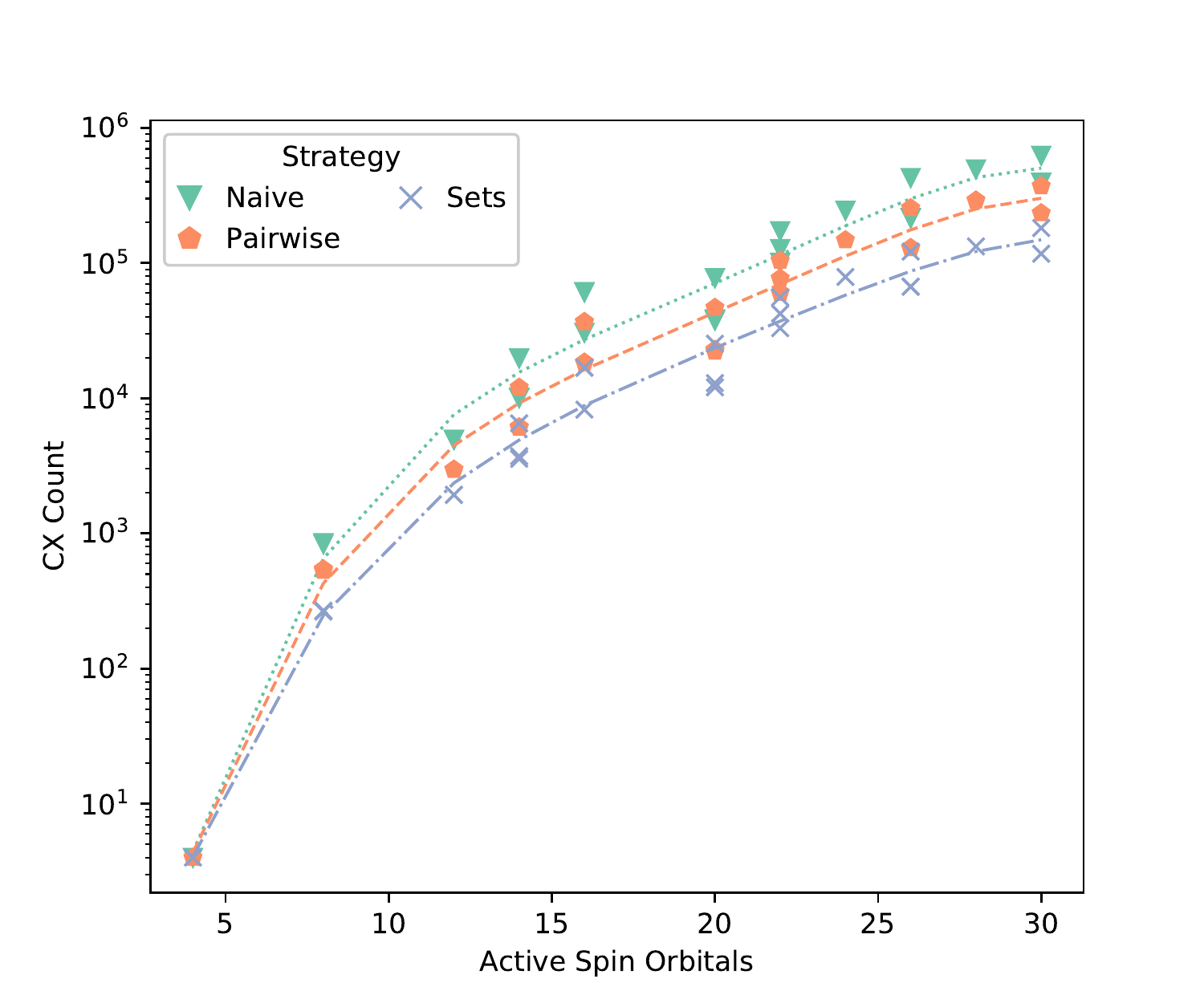} &
\includegraphics[scale=0.5]{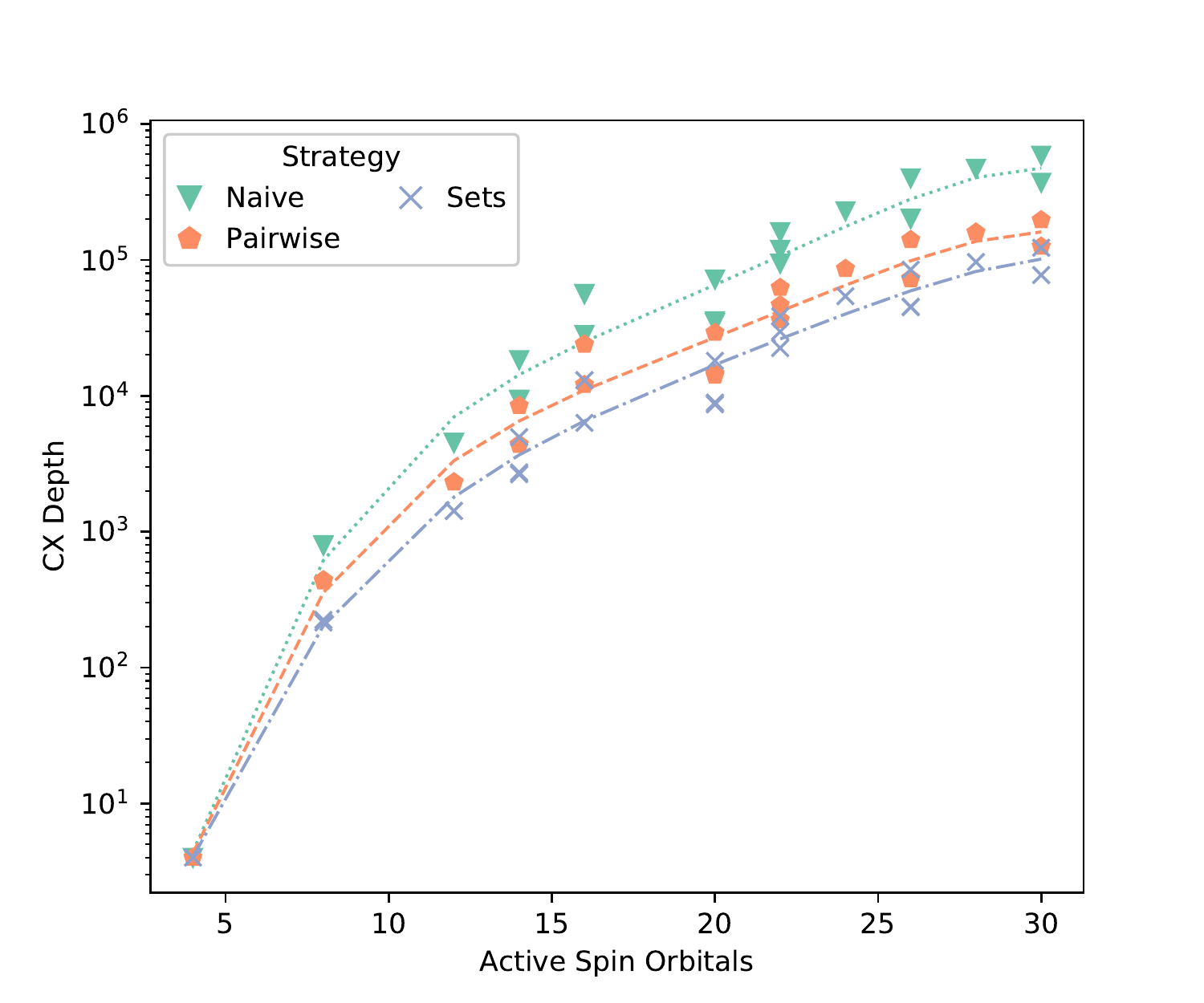}
\end{tabular}
\caption{Parity qubit encoding.}
\end{subfigure}

\caption{Comparison of compilation strategies for molecules with
varying active spin orbital counts using different qubit encoding
methods. A 4th-degree polynomial least-squares fit has been added to
suggest scaling.}
\label{fig:main_results}
\end{figure}

A comparison of \CX metrics for different compilation strategies,
active spin orbitals and qubit encoding methods is shown in
Figure~\ref{fig:main_results}.

The set-based synthesis strategy outperforms pairwise and naive
strategies on all encodings, but is on average outperformed by the
TLOS method for the JW encoding, particularly for larger systems with
more active spin orbitals.

\begin{figure}[tbh]
\centering

\begin{subfigure}[b]{\textwidth}
\begin{tabular}{l|c|c}
      & Mean \CX count reduction (\%) & Mean \CX depth reduction (\%)
      \\ \hline
      Pairwise Synthesis     & $40.0$  & $56.9$ \\
      Set-based Synthesis & $63.6$  & $71.9$ \\ \hline
\end{tabular}
\caption{Bravyi-Kitaev qubit encoding.}
\end{subfigure}
\vspace{3.5mm}

\begin{subfigure}[b]{\textwidth}
\begin{tabular}{l|c|c}
      & Mean \CX count reduction (\%) & Mean \CX depth reduction (\%)  \\\hline
      Pairwise Synthesis     & $49.9$  & $67.9$ \\
      Set-based Synthesis & $78.0$  & $82.1$ \\
      TLOS Synthesis & $82.6$ & $84.5$ \\\hline
\end{tabular}
\caption{Jordan-Wigner qubit encoding.}
\end{subfigure}
\vspace{3.5mm}

\begin{subfigure}[b]{\textwidth}
\begin{tabular}{l|c|c}
      & Mean \CX count reduction (\%) & Mean \CX depth reduction (\%)  \\\hline
      Pairwise Synthesis     & $38.1$  & $55.7$ \\
      Set-based Synthesis & $65.3$  & $72.1$ \\\hline
\end{tabular}
\caption{Parity qubit encoding.}
\end{subfigure}
\vspace{3.5mm}

\begin{subfigure}[b]{\textwidth}
\begin{tabular}{l|c|c}
      & Mean \CX count reduction (\%) & Mean \CX depth reduction (\%)  \\\hline
      Pairwise Synthesis     & $42.7$  & $60.2$ \\
      Set-based Synthesis & $69.0$  & $75.4$ \\\hline
\end{tabular}
\caption{All encodings. }
\end{subfigure}

\caption{Mean \CX metric reductions. All reductions are measured against the naive decomposition method. }
\label{fig:mean_results}
\figureline
\end{figure}

Set-based synthesis gives greater fractional reductions for larger
circuits than for smaller ones. For the largest circuits, up to 89.9\%
\CX depth reduction can be achieved, compared to the mean \CX
depth reduction of 75.4\% shown in Figure~\ref{fig:mean_results}. As
the compilation strategy is composed of several heuristics in
sequence, we do not at this stage argue that the asymptotic complexity
of the UCC ansatz can be reduced - in order to do this, we would need
to prove sufficient bounds on the size of sets found by graph
colouring, the \CX complexities of Clifford circuits required for
diagonalisation and the number of \CX gates produced by phase
polynomial synthesis.

\noindent
\textbf{Remark:\ } This compilation strategy assumes that the qubits
have all-to-all connectivity, so \CX gates are allowed between any two
qubits. When connectivity is constrained, \textit{routing} is required
to ensure the circuit conforms to the constraints. The typical
approach to this problem is SWAP network insertion
\cite{Childs:2019aa,Zulehner:2017aa,Lao:2019aa,Alexander-Cowtan:2019aa}.

\noindent
\textbf{Remark:\ } While VQE using the UCC ansatz is a candidate for
quantum advantage, there are no complexity-theoretic guarantees of
success. Should the advantage be sufficiently small, the low-degree
polynomial compilation time required for this strategy could be too
slow. In this case, we emphasise that \textit{co-design} of a
compilation strategy with the qubit encoding can give large
reductions, shown by the TLOS method, while reducing compilation time.

\section{Conclusions and Future Work}
\label{sec:conc}
The primary contribution of our paper is an empirically successful
method to efficiently synthesise the UCC ansatz to one- and two-qubit
gates. We have shown large average reductions in \CX metrics for the
Bravyi-Kitaev, Jordan-Wigner, and parity qubit encodings; although
alternative methods are competitive with ours for the JW encoding, we
emphasise that our strategy is valid for any other qubit encodings
which generate similar Trotterized excitation operators.  We note that
the reductions for the JW encoding are the greatest, with respect to
both metrics and both the pairwise and set-based synthesis methods.
This may suggest that this encoding has more exploitable redundancy
then the BK or P encodings.

We briefly discuss four future directions to explore.

\subsection{Applications to Measurement Reduction}
Measurement reduction for VQE is a method to simultaneously measure
terms in a Hamiltonian which commute, and therefore reduce the number
of circuits required to run \cite{jena2019pauli,
crawford2019efficient, zhao2019measurement}. For realistic devices,
assuming that the only available native measurements are single-qubit
$Z$-basis measurements, generating a Clifford circuit to diagonalise
this set is required. Minimising this Clifford circuit using
applications of Theorem~\ref{thm:pauli-chain} can reduce the \CX
overhead required for measurement reduction.

\subsection{Architecture-Aware Synthesis}
Instead of introducing a SWAP network to enforce connectivity
constraints on NISQ devices, recent work has explored the possibility
of resynthesising the circuit in a topologically aware manner, for
limited gate sets \cite{Kissinger:2019ac, Nash:2019aa,
wu2019optimization}. This constrained synthesis has been found to
typically produce lower \CX counts than SWAP networks, and phase
polynomials are a viable class of circuit for constrained synthesis
\cite{amy2019staq, Arianne-Meijer-van-de-Griend:2020aa}. If topologically
constrained phase polynomials can be composed with Clifford regions in
a manner that respects architecture, this would appear to be a
suitable strategy for those devices with limited connectivity.

\subsection{Applications to QAOA}
The Quantum Approximate Optimisation Algorithm (QAOA)
\cite{Farhi:2014aa} for combinatorial optimisation problems consists
of repeated blocks of `mixing' and `driver' exponentiated
Hamiltonians. The driver Hamiltonians are already diagonal, as they
encode a classical system, and typically the mixing Hamiltonians
correspond to single-qubit gates only. However, recent work on a
so-called Quantum Alternating Operator Ansatz \cite{Hadfield_2019}
introduces more complicated mixing Hamiltonians. These mixing
Hamiltonians could be amenable to our compilation strategy.

\subsection{Applications to Fault Tolerant Computation}
\label{sec:appftqc}
While this strategy was designed specifically for VQE, it can be
directly ported over to non-variational quantum algorithms for
Hamiltonian dynamics which require gate counts and qubit numbers too
high for NISQ computers. For the case where the Hamiltonian evolution
is approximated using product formulae, Gui et al.~\cite{gui2020term}
and van den Berg \& Temme~\cite{Berg:2020aa} have performed term
sequencing similar to our work in Section~\ref{sec:termseq} for
digital quantum simulation, in which a quantum evolution defined by a
time-dependent Hamiltonian is mapped to a quantum circuit. Reducing
Trotter error is more important for fault-tolerant algorithms than for
VQE, as it is the only significant non-correctable source of error,
and Gui et al.~argue that our term sequencing method would also
minimise Trotter error.

The efficacy of our proposed compilation strategy is greatly dependent
on the model of fault-tolerant computation. For example, in the model
presented by Litinski \cite{Litinski2019gameofsurfacecodes}, all
non-Clifford Pauli exponentials are performed natively by performing
ancilla measurements. In this model, the exponentials do not need to
be converted to one- and two-qubit gates at all.

Even for models which perform individual gates explicitly, our
proposed compilation strategy is optimised to reduce two-qubit gate
count and depth, which are not considered as important on planned
fault tolerant devices as non-Clifford gates, as the \CX gate can be
performed without magic state distillation. However, on surface
code-based computers performing lattice surgery, a two-qubit gate
between distant logical qubits can be as costly in the worst case as
magic state distillation \cite{Litinski_2019}; in general, two-qubit
gates increase the overhead from routing on surface codes. Therefore,
two-qubit gate reduction may still be a valuable optimisation.

Moreover, the circuits produced by the strategy are structured such
that all non-Clifford rotations reside in partitioned phase
polynomials, and will be approximated with $T$ gates and one-qubit
Cliffords. $T$-count and $T$-depth optimisation has been successfully
performed using the phase polynomial formalism via matroid
partitioning \cite{Amy2014Polynomial-Time}. The $T$-count of phase
polynomials generated via diagonalisation cannot be optimised using
phase-folding, as the parities are guaranteed to be unique, but
$T$-depth reduction could still be enabled by our strategy.

\section*{Acknowledgements}
The authors would like to thank John van de Wetering, Arianne Meijer,
David Zsolt-Manrique and Irfan Khan for helpful discussions, and
Matthew Amy for correspondence on the GraySynth algorithm.

\small
\bibliography{all}


\clearpage
\normalsize


\appendix

\section{Trotter Error}\label{app:trot-error}

Given the Lie-Trotter formula for our parameterised operator:

\begin{equation}
U_{\rho}(\vec{t}) = (\prod_{j}e^{\frac{t_j}{\rho}(\tau_j-\tau_j^{\dagger})})^{\rho}
\end{equation}

The Trotter error bound for this expansion, $\delta_{\rho}$, is
described in Low et al. \cite{low2019wellconditioned}:

\begin{equation}
\|\delta_{\rho}\| = \mathcal{O} \Big( ( \frac{1}{\rho}\sum_j \|t_j(\tau_j-\tau_j^{\dagger}) \| )^2 \Big) 
\end{equation}

Given a well-chosen reference state with a large overlap with the
exact wavefunction, low amplitudes will parameterise the ansatz, i.e.
$\forall j$, $t_j << 1$ \cite{Romero_2018}. Therefore, the
$\|t_j(\tau_j-\tau_j^{\dagger}) \|$ term will be small, particularly
when compared to current two-qubit gate errors of the order of $0.1\%$
or greater \cite{Wright:2019aa, Arute:2019aa}.

Conveniently, Gui et al. give evidence to suggest that term grouping
by mutual commutation such as we describe in this paper minimises
Trotter error compared to other methods \cite{gui2020term}. For a
rigorous theory of Trotter error, see Childs et al.
\cite{childs2019theory}.

\section{Phase Gadgets}\label{app:phase-gadgets}

\begin{theorem}
  We have the following laws for decomposition, commutation, and
  fusion of phase gadgets \cite{Cowtan:2019aa}.
  \begin{align*}
{\inltf{PhaseGadgetCNOT-lhs}} & \quad = \quad
{\inltf{PhaseGadgetCNOT}}   \\ \\
{\input{PhaseGadgetCommute0.tikz}} & \quad = \quad
{\input{PhaseGadgetCommute1.tikz}} \\ \\
{\input{PhaseGadgetFusion0.tikz}} & \quad = \quad
{\input{PhaseGadgetFusion1.tikz}}
  \end{align*}
\end{theorem}

\begin{figure}[th!]
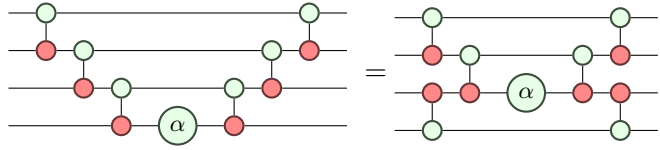

\begin{equation*}
  {\inltf{PhaseGadgetLadder}} = {\inltf{PhaseGadgetTree}}
\end{equation*}
\caption{Comparing worst-case and best-case patterns for constructing
phase gadgets with respect to \CX depth. The left shows a \CX ladder,
with a linear \CX depth, and the right shows the balanced-tree form,
with a logarithmic \CX depth.}
\label{fig:ladder_vs_tree}
\end{figure}

Phase gadgets are invariant under qubit permutation. For an $n$-qubit
phase gadget, this gives a choice of $C_{n-1}n!$ different \CX
arrangements, where $C_n$ is the $n$-th Catalan number.
Figure~\ref{fig:ladder_vs_tree} shows example arrangements.

\section{Clifford Gates and Pauli Gadgets}\label{app:cliff-commutation}

\begin{theorem}
\label{fig:exponential_clifford_rules}
We have the following laws for commuting single-qubit Clifford gates
through Pauli gadgets \cite{Cowtan:2019aa}.
\begin{figure}[ht!]
\begin{subfigure}[b]{0.5\textwidth}
\begin{equation*}
{\input{PauliExpZZa.tikz}} = {\input{PauliExpZZb.tikz}}
\end{equation*}
\begin{equation*}
{\input{PauliExpZXa.tikz}} = {\input{PauliExpZXb.tikz}}
\end{equation*}
\begin{equation*}
{\input{PauliExpZYa.tikz}} = {\input{PauliExpZYb.tikz}}
\end{equation*}
\end{subfigure}
\begin{subfigure}[b]{0.5\textwidth}
\begin{equation*}
{\input{PauliExpXZa.tikz}} = {\input{PauliExpXZb.tikz}}
\end{equation*}
\begin{equation*}
{\input{PauliExpXXa.tikz}} = {\input{PauliExpXXb.tikz}}
\end{equation*}
\begin{equation*}
{\input{PauliExpXYa.tikz}} = {\input{PauliExpXYb.tikz}}
\end{equation*}
\end{subfigure}
\begin{equation*}
{\input{PauliExpHZa.tikz}} = {\input{PauliExpHZb.tikz}}
\end{equation*}
\begin{equation*}
{\input{PauliExpHXa.tikz}} = {\input{PauliExpHXb.tikz}}
\end{equation*}
\begin{equation*}
{\input{PauliExpHYa.tikz}} = {\input{PauliExpHYb.tikz}}
\end{equation*}
\begin{subfigure}[b]{0.5\textwidth}
\begin{equation*}
{\input{PauliExpSZa.tikz}} = {\input{PauliExpSZb.tikz}}
\end{equation*}
\begin{equation*}
{\input{PauliExpSXa.tikz}} = {\input{PauliExpSXb.tikz}}
\end{equation*}
\begin{equation*}
{\input{PauliExpSYa.tikz}} = {\input{PauliExpSYb.tikz}}
\end{equation*}
\end{subfigure}
\begin{subfigure}[b]{0.5\textwidth}
\begin{equation*}
{\input{PauliExpVZa.tikz}} = {\input{PauliExpVZb.tikz}}
\end{equation*}
\begin{equation*}
{\input{PauliExpVXa.tikz}} = {\input{PauliExpVXb.tikz}}
\end{equation*}
\begin{equation*}
{\input{PauliExpVYa.tikz}} = {\input{PauliExpVYb.tikz}}
\end{equation*}
\end{subfigure}
\end{figure}
\end{theorem}

\begin{theorem}
\label{fig:exponential_cx_rules}
We have the following laws for commuting \CX gates through Pauli
gadgets \cite{Cowtan:2019aa}.
\begin{figure}[ht!]
\begin{subfigure}[b]{0.5\textwidth}
\begin{equation*}
{\input{PauliExpCXZra.tikz}} = {\input{PauliExpCXZrb.tikz}}
\end{equation*}
\begin{equation*}
{\input{PauliExpCXXra.tikz}} = {\input{PauliExpCXXrb.tikz}}
\end{equation*}
\begin{equation*}
{\input{PauliExpCXYra.tikz}} = {\input{PauliExpCXYrb.tikz}}
\end{equation*}
\end{subfigure}
\begin{subfigure}[b]{0.5\textwidth}
\begin{equation*}
{\input{PauliExpCX_ZIa.tikz}} = {\input{PauliExpCX_ZIb.tikz}}
\end{equation*}
\begin{equation*}
{\input{PauliExpCX_XIa.tikz}} = {\input{PauliExpCX_XIb.tikz}}
\end{equation*}
\begin{equation*}
{\input{PauliExpCX_YIa.tikz}} = {\input{PauliExpCX_YIb.tikz}}
\end{equation*}
\end{subfigure}
\par\bigskip
\begin{equation*}
{\input{PauliExpCX_ZXa.tikz}} = {\input{PauliExpCX_ZXb.tikz}}
\end{equation*}
\begin{equation*}
{\input{PauliExpCX_XZa.tikz}} = {\input{PauliExpCX_XZb.tikz}}
\end{equation*}
\begin{equation*}
{\input{PauliExpCX_XYa.tikz}} = {\input{PauliExpCX_XYb.tikz}}
\end{equation*}
\end{figure}
\end{theorem}

\section{Proof of Corollary 5.4}
\label{app:corollary54proof}

All commuting sets of $m$ Pauli gadgets over $n$ qubits are
diagonalisable using Theorem~\ref{thm:pauli-chain} if all sets of $m$
pairs of Paulis are either compatible with
Theorem~\ref{thm:pauli-chain} or already contain a diagonal qubit. We
prove by enumerating over these sets of pairs that this compatibility
is satisfied for the case $m = 3$, and therefore for $m < 4$.
Compatibility is not satisfied for $m = 4$, and therefore any $m > 3$.

A short script to verify this can be found at
\url{https://github.com/CQCL/tket_benchmarking/blob/master/compilation_strategy/corollaries/corollary54.py}.

\section{Proof of Corollary 5.5}
\label{app:corollary55proof}
Enumerating over all commuting sets of gadgets over 4 qubits and
finding at least one pair of compatible qubits for each commuting set
is sufficient proof, as each commuting set of gadgets over fewer than
4 qubits is just a special case of a 4-qubit set. However, each unique
commuting set is defined by a Clifford circuit. There are more than
$4.7 \times 10^{10}$ Clifford circuits over 4 qubits, ignoring global
phase \cite{PhysRevA.88.052307}. As an optimisation, we instead search
over all the \textit{generators} of each commuting set of gadgets.
Each commuting set over 4 qubits can be generated by taking products
from a commuting set of 4 Pauli strings. It is therefore sufficient to
find at least one pair of compatible qubits for each commuting set of
4 Pauli strings.

A short script to verify this can be found at
\url{https://github.com/CQCL/tket_benchmarking/blob/master/compilation_strategy/corollaries/corollary55.py}.

\section{Operator generation}
\label{app:qboperator}
For the Jordan-Wigner and Bravyi-Kitaev encodings, \tket
\texttt{QubitPauliOperator} objects were produced using EUMEN, an
under-construction software platform for quantum chemistry on quantum
computers. Excitation operators $\tau_j$ are calculated from the
molecules' spin orbitals, after which they are converted into
\texttt{QubitPauliOperator} objects, which represent the $U(\vec{t})$
operators from Equation~\ref{eq:sum-expand}. These objects contain a
python dictionary from Pauli string to symbolic expression
representing $t^{'}_j$. The coefficients $a_j$ are dependent on the
molecular geometry, and are unimportant to the compilation strategy.

The qubit operators for the parity encoding were obtained from Qiskit
\cite{Qiskit}, and converted to \tket native
\texttt{QubitPauliOperator} objects.

All \texttt{QubitPauliOperator} objects are serialised and stored at
\url{https://github.com/CQCL/tket_benchmarking/tree/master/compilation_strategy/operators}.

For the templated lexicographical operator sequence (TLOS) method, the
circuits are generated from the excitation operators, rather than a
dictionary of Pauli strings to expressions, and therefore bypass the
$U(\vec{t})$ operators stage of Equation~\ref{eq:sum-expand} entirely.
Rather than serialising the corresponding operators, the relevant TLOS
circuits are stored in the OpenQASM format
\cite{Cross2017Open-Quantum-As} at
\url{https://github.com/CQCL/tket_benchmarking/tree/master/compilation_strategy/TLOS_qasm_files}.

We do not include the operations required to generate a chosen
reference state $\ket{\Phi_0}$, as they are irrelevant to the
strategy.

\end{document}